\documentclass[12pt]{article}

\usepackage[utf8]{inputenc}
\usepackage[T1]{fontenc}
\usepackage{lmodern}
\usepackage{times}
\usepackage{graphicx}
\usepackage{amsmath,amssymb}
\usepackage{geometry}
\usepackage{hyperref}
\usepackage{cite}
\usepackage{setspace}
\usepackage{url}
\usepackage{enumitem}
\usepackage{authblk}

\newcommand\arcsec{\hbox{$^{\prime\prime}$}}

\title{Warm Gas in the Vicinity of a Supermassive Black Hole 13 Billion Years Ago}
\author[1,2]{K. Tadaki}
\author[3,4]{F. Esposito}
\author[4]{L. Vallini}
\author[5,6]{T. Tsukui}
\author[7]{T. Saito}
\author[2,8]{D. Iono}
\author[9]{T. Michiyama}
\affil[1]{{\footnotesize Faculty of Engineering, Hokkai-Gakuen University, Toyohira-ku, Sapporo 062-8605, Japan}}
\affil[2]{{\footnotesize National Astronomical Observatory of Japan, 2-21-1 Osawa, Mitaka, Tokyo 181-8588, Japan}}
\affil[3]{{\footnotesize Dipartimento di Fisica e Astronomia, Università degli Studi di Bologna, Via P. Gobetti 93/2, I-40129 Bologna, Italy}}
\affil[4]{{\footnotesize Osservatorio di Astrofisica e Scienza dello Spazio (INAF--OAS), Via P. Gobetti 93/3, 40129 Bologna, Italy}}
\affil[5]{{\footnotesize Research School of Astronomy and Astrophysics, Australian National University, Cotter Road, Weston Creek, ACT 2611, Australia}}
\affil[6]{{\footnotesize ARC Centre of Excellence for All Sky Astrophysics in 3 Dimensions (ASTRO 3D), Australia}}
\affil[7]{{\footnotesize Faculty of Global Interdisciplinary Science and Innovation, Shizuoka University, 836 Ohya, Suruga-ku, Shizuoka 422-8529, Japan}}
\affil[8]{{\footnotesize Department of Astronomical Science, SOKENDAI (The Graduate University for Advanced Studies), 2-21-1 Osawa, Mitaka, Tokyo 181-8588, Japan}}
\affil[9]{{\footnotesize Faculty of Information Science, Shunan University, 843-4-2, Gakuendai, Shunan, Yamaguchi 745-8566, Japan}}
\date{}

\begin{document}

\maketitle

\begin{abstract}
\begin{sloppypar}
\begin{spacing}{1.2}
{\large
Quasars, powered by supermassive black holes (SMBH), are among the brightest objects in the universe. In the vicinity of an SMBH, X-ray photons from an active galactic nucleus (AGN) can heat the surrounding gas to several hundred kelvin. Here we report observations of dust continuum and CO $J=13-12$ and $J=14-13$ line emissions at a resolution of 130 parsecs in a luminous quasar at $z=6$. We successfully detected these high-$J$ CO line emissions from warm gas in a compact disk component. 
The CO luminosity ratio in the central region of the compact disk is consistent with theoretical models in which X-ray heating dominates the CO excitation and the gas column density is as high as $10^{25}\,\mathrm{cm}^{-2}$. 
This demonstrates that high-resolution observations of high-$J$ CO lines are promising ways to identify extremely dust-obscured quasars in the early universe. 
}
\end{spacing}
\end{sloppypar}
\end{abstract}

\clearpage

In the universe at $z>6$, corresponding to 13 billion years ago, luminous quasars containing a SMBH with a mass exceeding $10^9\,M_\odot$ have been discovered \cite{Mortlock2011,Matsuoka2018,Onoue2019,Yang2021}. Many of these quasars were originally identified by wide-field surveys at optical and near-infrared wavelengths, and were subsequently detected at submillimeter wavelengths, with some reaching total infrared luminosities of $L_{\rm IR}\sim10^{13}\,L_\odot$ \cite{Bertoldi2003,Wang2013}. Studying the interstellar medium surrounding AGN in such luminous quasars is crucial to understanding the interaction between the AGN and its surrounding environment. Rotational-vibrational excitation lines of molecular hydrogen in the near-infrared are useful for tracing warm (1000-2000 K) molecular gas directly affected by an AGN \cite{Davies2014}. However, it is difficult to observe these emission lines in the central regions of distant quasars due to sensitivity and spatial resolution limitations, even with JWST. Observations of highly excited rotational transitions of carbon monoxide (high-$J$ CO lines) are an alternative approach to probe warm gas in quasars \cite{Gallerani2014,Carniani2019,Li2020,Pensabene2021}. Previous observations of high-$J$ CO lines had low spatial resolution and could not constrain the location of the warm gas and its heating mechanism. The scope of this paper is to understand the relationship between warm gas and AGN in the central regions of quasars in the early universe.

\subsection*{ALMA Observations}

To directly probe the warm gas in the vicinity of an AGN, we have observed CO $J=13-12$ and CO $J=14-13$ lines with upper energy levels of $E_u/k_B=500-600$ K, as well as dust continuum emission, in J231038.88+185519.7 (hereafter J2310+1855) at $z=6.00$ with a high spatial resolution of 130 pc (0.022\arcsec) using the Atacama Large Millimeter/submillimeter Array (ALMA) (Fig.~\ref{tab:tab1}). J2310+1855 is one of the brightest quasars in both optical and submillimeter emission \cite{Wang2013,Jiang2016}. The detection of a broad Mg\,\textsc{ii} emission line extending over 3000 km\,s$^{-1}$ implies a SMBH with a mass of $M_{\rm BH}=(2-4)\times10^9\,M_\odot$ in the center of J2310+1855 \cite{Farina2022} (Fig.~\ref{fig:ex-fig1}). Figure~\ref{fig:fig1}a shows the ALMA image of the 1.4 mm continuum emission at a resolution of 130 pc (Fig.~\ref{tab:tab2}). We found that the dust emission comes from a compact disk with an effective radius of $R_e=266\pm14$ pc (Methods and Fig.~\ref{fig:ex-fig2}). Such a centrally concentrated gas disk may have been formed by efficient gas transport driven by a major merger \cite{Xu2014,Xu2015}, and its detection has been reported in other luminous quasars at $z\sim6-7$ \cite{Venemans2019,Walter2022,Meyer2023,Neeleman2023}. High resolution observations of [C\,\textsc{ii}] line emission, which traces cold gas, have been used extensively to study the distribution of molecular gas masses and black hole masses in the central compact disk of quasars \cite{Walter2022,Meyer2023}. In contrast, warm gas at these spatial scales has received comparatively little attention and remains largely unexplored. We have now detected, for the first time, the CO $J=13-12$ and CO $J=14-13$ emissions in the central compact disk using ALMA images with 250 pc resolution (Fig.\ref{fig:fig1}b,c).

Near an AGN, X-rays from an accretion disk and its associated coronal gas are expected to heat the molecular gas to high temperatures, resulting in bright high-$J$ CO line emission. This is in contrast to star forming regions, where cold gas dominates. Even higher-$J$ CO emission has been detected in other quasars at $z>6$ \cite{Gallerani2014,Pensabene2021}, suggesting a strong contribution from X-ray dominated regions (XDRs). In the nearest quasar, Mrk 231 at $z=0.04$, CO $J=13-12$ line emission has been detected by the Herschel Space Observatory \cite{vanderWer2010}. However, due to the low spatial resolution of 15 kpc, there is still no direct evidence that the high-$J$ CO emission is associated with warm gas in the vicinity of an AGN. High-$J$ CO lines from high-redshift quasars are shifted to wavelengths with good atmospheric transmission (1--2 mm) and can be observed with large ground-based telescopes that provide high spatial resolution data. Our high-resolution ALMA images directly indicate the presence of warm gas in the central region where an AGN is expected to be located. 

\subsection*{Comparisons with Theoretical Models}

\begin{figure}[htbp]
  \centering
  \includegraphics[width=1.0\textwidth]{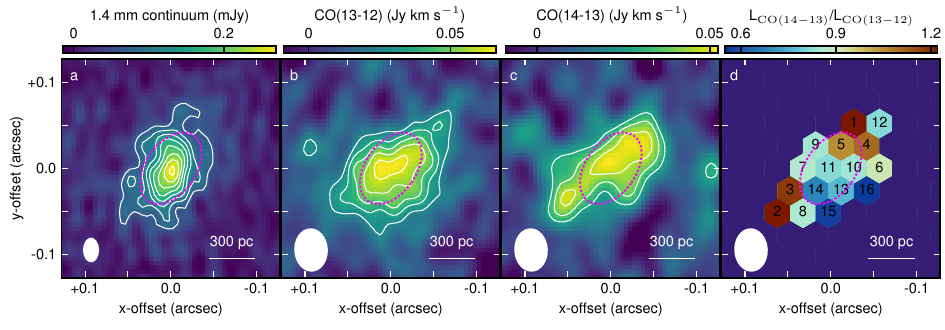}
  \caption{
  Morphology of a luminous quasar at $z=6.0$, J2310+1855. ALMA maps of (a) 1.4 mm continuum, (b) CO $J=13-12$, (c) CO $J=14-13$ emission, and (d) CO luminosity ratio. The contours are plotted every 1$\sigma$ from 3$\sigma$ in a and b and every 2$\sigma$ from 5$\sigma$ in c. The angular resolution is 0.029\arcsec$\times$0.019\arcsec~ (a position angle of 0.5 degree) for the continuum and 0.051\arcsec$\times$0.039\arcsec~ (3.1 degree) for the CO maps. The magenta dashed line shows the effective radius of the compact disk ($R_e=266$ pc).}
  \label{fig:fig1}
\end{figure}

The observed luminosity ratios of CO $J=14-13$ and CO $J=13-12$ emissions range from $L_{\rm CO(14-13)}/L_{\rm CO(13-12)} = 0.5\text{--}1.4$ (Fig.~\ref{fig:fig1}d). We then identify the primary heating source of the warm gas by comparing the CO ratios with theoretical models, that consider three physical mechanisms: heating associated with star formation activity in photodissociation regions (PDRs), heating associated with black holes in XDRs, and heating associated with shock waves. Previous Herschel observations have shown that the luminosity ratios of galaxy-integrated CO line emission do not necessarily provide an accurate classification of the heating mechanisms in the interstellar medium \cite{Mashian2015}. For example, although the nearby ULIRG NGC 6240 has a CO spectral line energy distribution (SLED) similar to that of Mrk 231, its detailed studies suggest that the gas is heated by shocks \cite{Meijerink2013}. We address this issue of the heating mechanism by comparing gas and dust data at smaller spatial scales, which are not available from observations of nearby galaxies, with physically motivated models \cite{Vallini2019,Esposito2024} (Methods).

In the PDR models, where far ultraviolet (6--13.6 eV) photons from massive OB-type stars regulate the heating and ionization of the gas, CO ratios increase with star formation rates. The total star formation rate of J2310+1855 is SFR=$1000-3000~M_\odot$\,yr$^{-1}$, including the host galaxy \cite{Shao2019,Tripodi2022}. Even taking into account the extreme stellar radiation, which is $10^6$ times larger than the far ultraviolet radiation field in the Milky Way ($G_0=10^6$), the CO ratios above 0.5 cannot be explained by the PDR model (Fig.~\ref{fig:fig2}). Theoretically, high-$J$ CO emission can be excited even in PDRs if the gas disk consists entirely of clumps with densities of $10^6$\,cm$^{-3}$ \cite{Esposito2022}. However, since such high-density regions are expected to have a low volume filling factor, our model assumes a lognormal distribution of the gas densities of clumps \cite{Vallini2017}. Observations of local AGN host galaxies also support that PDRs contribute little to the excitation of such high-$J$ CO emission lines \cite{Esposito2024,Mingozzi2018}.

Next, we consider the XDR heating. Since J2310+1855 is detected in X-rays with a luminosity of $L_{2-10\,\mathrm{keV}}=6.9\times10^{44}$ erg\,s$^{-1}$ \cite{Vito2019}, it is reasonable to assume that the central X-ray source has a strong influence on the excitation conditions of the molecular gas in the central region. We assume that the gas in the compact disk with $R_e=266$ pc is primarily responsible for the attenuation of the X-ray emission from an AGN. In regions radially away from the center, the gas column density accumulated along the radial direction increases, leading to a rapid decrease in the incident X-ray flux due to a shielding effect (Fig.~\ref{fig:ex-fig3}). We compare the observed luminosity ratios of the CO $J=14-13$ and CO $J=13-12$ emissions with those in the XDR models of the compact disk with gas column densities of $N_H=10^{24.5}$--$10^{25.5}$ cm$^{-2}$ at a galactocentric radius of 1 kpc (Fig.~\ref{fig:fig2}). The XDR models with larger column densities have smaller CO ratios at the same radius. The XDR models then reasonably explain a high CO ratio of $L_{\rm CO(14-13)}/L_{\rm CO(13-12)} \sim~0.8$ at radii of $R<200$ pc and a decreasing trend to a lower $L_{\rm CO(14-13)}/L_{\rm CO(13-12)}\sim~0.5$ in the southwestern region (ID 15,16 in Fig.~\ref{fig:fig1}d). Note that the gas column density here is measured in the radial direction of the compact disk, not in the line of sight. X-ray observations of several high-$z$ quasars, including J2310+1855, suggest that the gas column density in the line-of-sight direction is typically $N_H<9\times10^{22}$ cm$^{-2}$ \cite{Vito2019}, which is much smaller than what we consider. 

In the northwestern and southeastern regions of the compact disk (ID 1--4 in Fig.~\ref{fig:fig1}d), the CO $J=14-13$ emission is almost thermalized ($L_{\rm CO(14-13)}/L_{\rm CO(13-12)}\sim1.25$) and cannot be explained by either the PDR or the XDR model. We consider the possibility of mechanical heating by shocks. Powerful outflows are often detected in luminous quasars \cite{Bischetti2019,Izumi2021}. The interaction of quasar-driven outflows with the surrounding gas is expected to produce a shock wave that heats the gas to a high temperature. If the direction of bipolar outflows is inclined relative to a disk plane, the shock heating would affect gas in regions slightly away from the center. We use a planar shock model \cite{Flower2010,Flower2015} to predict CO fluxes in the large velocity gradient approximation. The model calculates temperature and density profiles taking into account several processes: gas-grain interactions, grain charge variation, and momentum transfer between charged and neutral fluids. CO excitation depends on a shock velocity and a gas density of the pre-shock gas. The shock model with a velocity of $v=20-40$ km s$^{-1}$ and a density of $n=10^5$ cm$^{-3}$ explains a high CO ratio of $L_{\rm CO(14-13)}/L_{\rm CO(13-12)}=1.0-1.1$, independent of galactocentric radii.

Since the gas in the central region can be excited by either XDR or shocks, it is difficult to distinguish between the two mechanisms based on the CO ratios alone. Therefore, we use the dust information from the ALMA observations. An AGN produces ultraviolet continuum emission as well as X-rays and efficiently heats the dust. In the nuclear region ($<$200 pc) of Mrk 231, the dust emission is likely to be dominated by a warm (100 K) component, rather than a cold (40 K) component \cite{Gonzalez-Alfonso2010}, supporting dust heating by an AGN. In contrast, shocks do not affect dust, while gas is compressed and heated to higher temperatures \cite{Meijerink2013}. We have detected multi-wavelength continuum emission in J2310+1855 with a resolution of 460 pc (0.08\arcsec) and with a resolution of 1.4 kpc (0.24\arcsec) (Fig.~\ref{fig:ex-fig4} and Fig.~\ref{tab:tab3}). By characterizing the spatially resolved spectral energy distributions (SEDs) of the dust emission (the left panel of Fig. ~\ref{fig:fig3}), we find that the dust temperature is $T_d=74^{+7}_{-5}$ K in the central region (within a radius of 230 pc) and $T_d=59^{+4}_{-4}$ K in the extended region (within a radius of 700 pc), which is consistent with a previous study \cite{Shao2022}. In a quasar at $z = 4.4$, the dust temperature has been reported to increase towards smaller galactocentric radii, reaching up to 60--70 K within the central 500 pc \cite{Tsukui2023}. Furthermore, in the central 110 pc region of a quasar at $z = 6.9$, the dust temperature is even higher, exceeding 132 K \cite{Walter2022}. Since that shocks do not heat the dust, measuring the dust temperature in the small central regions of quasars using high spatial resolution observations is a powerful way to prove that the dust and gas are being heated by the AGN. The higher dust temperature in the central region supports the scenario that the gas in the central region is heated by XDR rather than by shocks.

In addition to the CO $J=13-12$ and CO $J=14-13$ lines, we have detected CO $J=6-5$ and CO $J=8-7$ emission lines with an upper energy level of $E_u/k_B=100-200$ K at a spatial resolution of 460 pc (Fig.~\ref{fig:ex-fig5} and Fig.~\ref{tab:tab3}). A CO SLED including mid-$J$ CO lines is helpful for understanding the heating mechanism acting on molecular gas. Here we consider the effect of dust attenuation on CO SLEDs. The multi-wavelength analysis of the dust SED shows that the optical depth of the dust emission is unity at the rest-frame wavelength of $\lambda_0=266^{+52}_{-49}~\mu$m in the central region (the left panel of Fig.~\ref{fig:fig3}). In ISMs where gas and dust are well mixed, emissions at the shorter wavelengths, such as CO $J=13-12$ ($\lambda$=200 $\mu$m) and CO $J=14-13$ ($\lambda$=186 $\mu$m) lines, can be significantly attenuated by dust \cite{Rangwala2011}. We therefore correct the CO line fluxes for dust attenuation. The corrected CO SLED in the central region peaks at $J \geqq 9$ while the galaxy-integrated one peaks at $J=8$ \cite{Li2020} (the right panel of Fig.~\ref{fig:fig3}). This radial gradient of the CO excitation supports the scenario that the XDR heating preferentially works in the central region. We have integrated the XDR model out to a radius of 230 pc to compute a CO SLED normalized at $J=13$. The XDR model is broadly consistent with the CO SLED in the central region. On the other hand, the observations indicate that the CO lines are still highly excited in the outer region at galactocentric radii of $R=230-700$ pc, where the X-ray emission is attenuated by gas in the compact disk. Since neither PDR nor XDR can reproduce such a CO SLED, shocks are probably responsible for the CO excitation in the outer region as well as in the northwestern and southeastern regions of the compact disk (ID 1-4 in Fig.~\ref{fig:fig1}d).

\subsection*{Spatial Extent of Outflowing Gas}

We search for signatures of outflows in J2310+1855 based on ALMA observations of ionized carbon ([C\,\textsc{ii}]) emission to explore the physical origins of shocks. We use all archival data to obtain a deep spectrum of [C\,\textsc{ii}] emission with a spatial resolution of 8 kpc (Fig.~\ref{fig:fig4}). The spectrum of the whole galaxy is well characterized by two Gaussian components (the top panel of Fig.~\ref{fig:fig4}), which can be interpreted as gas in the rotating disk of the host galaxy. Previous studies of the gas kinematics of this quasar report a rotation velocity of 150--180 km s$^{-1}$ before inclination correction \cite{Shao2022,Tripodi2023}. After subtracting the disk component, we find excess emission at the 5$\sigma$ level in four continuous bins in the velocity range of $-465 < v < -345$ km\,s$^{-1}$ (the bottom panel of Fig.~\ref{fig:fig4}). Such a high-velocity wing component is independent of the rotating disk of the host galaxy and is likely to be an outflow component. The [C\,\textsc{ii}] flux of the blue wing component is $0.092\pm0.011$ Jy\,km\,s$^{-1}$, in broad agreement with a previous study \cite{Tripodi2022}. Assuming a circular Gaussian profile, we find that the wing component has a circularized effective radius of $R_{e,\rm circ}=2.4\pm0.4$ kpc kpc from the visibility fit to all data in the ALMA archive (Methods and  Fig.~\ref{fig:ex-fig6}). On the other hand, the disk component in the velocity range from $-315 < v < +315$ km\,s$^{-1}$ has $R_{e,\rm circ}=1.3\pm0.1$ kpc. Our size measurements indicate that the [C\,\textsc{ii}] emission of the wing component is distributed throughout the galaxy disk. The gas in the outer region could be highly excited by shocks generated by the interaction of the galaxy-scale outflow with the gas in the compact disk. Such interactions have also been suggested in a nearby AGN \cite{Saito2022} and a high redshift quasar \cite{Vayner2023}. 

\subsection*{An Approach to Searching for Dust-Obscured Quasars}

Figure~\ref{fig:fig5} shows the geometric configuration of the two disks and outflow components of J2310+1855 as inferred from the ALMA observations. Independent analyses of [CII] kinematics show that a gas disk is inclined by about 40 degrees \cite{Shao2022,Tripodi2023}. This is consistent with the large difference between the column density in the line of sight estimated from X-ray observations ($N_H < 9\times10^{22}$ cm$^{-2}$) \cite{Vito2019} and the column density in the radial direction of the compact disk estimated from the CO line ratio measurements ($N_H\sim10^{25}$ cm$^{-2}$). If the compact disk were observed from the radial direction, J2310+1855 would be classified as an extremely dust-obscured quasar with $N_H\sim10^{25}$ cm$^{-2}$ and would be missed in optical and X-ray observations. One of the scenarios for the growth of SMBHs is that gas-rich galaxies collide and merge, leading to an active phase of dusty star formation and a dusty AGN \cite{Hopkins2006}. During this phase, an SMBH is fed by gas at the galactic center. In nearby late-stage mergers, an AGN is encapsulated by obscuring material with a column density of $N_H>10^{23}$ cm$^{-2}$, which covers almost the entire circumference of the X-ray source \cite{Ricci2017}. In addition, about 30\% of AGNs detected with hard X-rays have a column density of $N_H>10^{24}$--$10^{25}$ cm$^{-2}$ \cite{Ricci2015}. The fraction of obscured AGN increases to 50\% at $z\sim3$ \cite{Lanzuisi2018}. Cosmological simulations also predict that there are more dust-obscured quasars in the early universe than known optically luminous quasars discovered to date \cite{Ni2020}. It is difficult to identify such dust-obscured quasars at high redshift in wide-field surveys at optical/near-infrared wavelengths where dust attenuates the emission. 

\begin{figure}[htbp]
  \centering
  \includegraphics[width=0.6\textwidth]{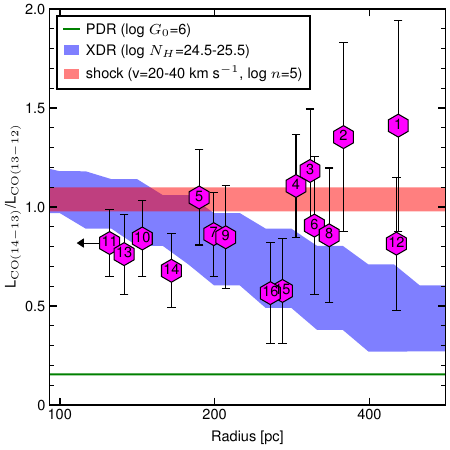}
  \caption{
  A variation of luminosity ratio between CO $J=14-13$ and CO $J=13-12$ emissions at different galactocentric radii. The magenta hexagons denote the CO ratios determined from ALMA observations, corresponding to the IDs used in Fig. \ref{fig:fig1}d. Theoretical model comparisons are shown with a green line representing PDR models with $G_0=10^6$, blue shaded regions representing XDR models with gas column densities ranging from $N_H=10^{24.5}$ cm$^{-2}$ to $N_H=10^{25.5}$ cm$^{-2}$, and red shaded regions representing shock models with velocities ranging from $v=20$ km s$^{-1}$ to $v=40$ km s$^{-1}$ and a gas density of $n=10^5$ cm$^{-3}$.
}
  \label{fig:fig2}
\end{figure}

High resolution observations of highly excited CO line emission could be useful to identify such dust-obscured quasars even with $N_H\sim10^{25}$ cm$^{-2}$. Figure~\ref{fig:fig6} shows a comparison of the galaxy-integrated CO SLED for J2310+1855 \cite{Carniani2019,Li2020} with other optically luminous quasars \cite{Pensabene2021,Riechers2009,Wang2019,Yang2019,Li2024,Xu2024} and dusty star-forming galaxies \cite{Riechers2013,Salome2012,Lee2021,Jarugula2021,Tsujita2022}. J2310+1855 is among the brightest unobscured quasars at $z=6-7$ in terms of CO luminosity. Although there is no significant difference in the shape of the CO SLED between J2310+1855 and CO-faint quasars, we cannot conclude from the observation of this single object that X-ray heated gas is present in the centers of all quasars. On the other hand, similarly bright high-J CO line emissions have been detected in dusty star-forming galaxies at $z=4-7$. The CO luminosity of these objects remains strong even at $J_\mathrm{up}=9-10$, suggesting the possible presence of an X-ray source such as an AGN at their centers. Hot dust-obscured galaxies with bolometric luminosities of $L_{\rm bol}>10^{14}\,L_\odot$ are also good candidate for extremely dust-obscured quasars, since their SEDs suggest the existence of hot dust with a temperature of about 450 K \cite{Tsai2015}. Future high-resolution observations with ALMA will allow us to probe the warm gas in the centers of dusty star-forming galaxies and hot dust-obscured galaxies and to study their excitation mechanisms. As shown in Figure~\ref{fig:fig2}, the measured luminosity ratio of $L_{\rm CO(14-13)}/L_{\rm CO(13-12)}$ allows us to rule out the possibility of gas heating by a starburst with high SFRs. These candidates of dust-obscured quasars may be at a different evolutionary stage than unobscured quasars where strong AGN-driven outflows have removed the surrounding obscuring material. In this case, only XDR heating would dominate the excitation of high-$J$ CO emission lines, so that, in contrast to J2310+1855, the CO line ratio would be expected to decrease monotonically from the center to the outer region. JWST observations are also finding a new population of less luminous AGN candidates with a dusty core at $z>4$ \cite{Matthee2024,Barro2024}. A complete study of quasars in different phases, including extremely dust-obscured ones, is important for a comprehensive understanding of their evolution.

\begin{figure}[htbp]
  \centering
  \includegraphics[width=1.0\textwidth]{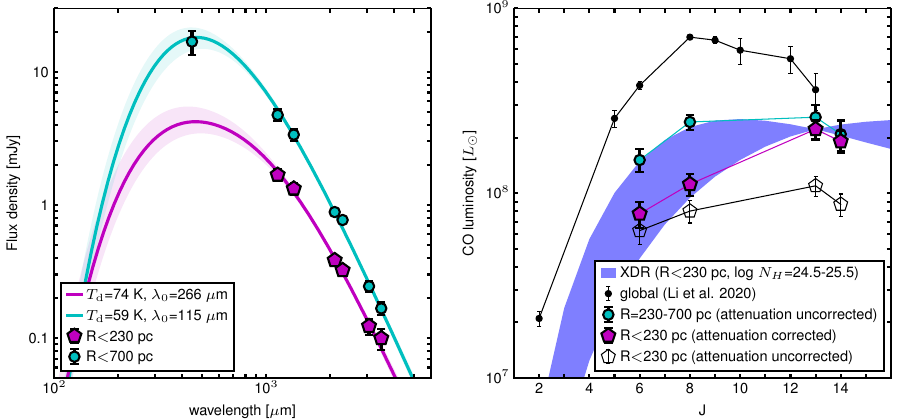}
  \caption{
Energy distributions of dust continuum and CO line emissions in J2310+1855. 
Left, dust SEDs in areas inside a radius of 230 pc (magenta pentagons) and inside a radius of 700 pc (cyan circles). Magenta and cyan lines indicate the best-fit models of modified blackbody radiation with $T_d$=74 K and $T_d$ =59 K, respectively. The shaded regions in magenta and cyan denote 1$\sigma$ uncertainties associated with these models. Right, pentagons, cyan circles and black circles represent CO SLEDs in the central region (R$<$230 pc), in the outer region (R=230-700 pc), and in the whole region \cite{Li2020}, respectively. Magenta pentagons show attenuation-corrected CO luminosities while open pentagons show uncorrected luminosities. A blue shaded region corresponds to the XDR model with gas column densities ranging from $N_H=10^{24.5}$ cm$^{-2}$ to $N_H=10^{25.5}$ cm$^{-2}$.
}
  \label{fig:fig3}
\end{figure}

\begin{figure}[htbp]
  \centering
  \includegraphics[width=0.6\textwidth]{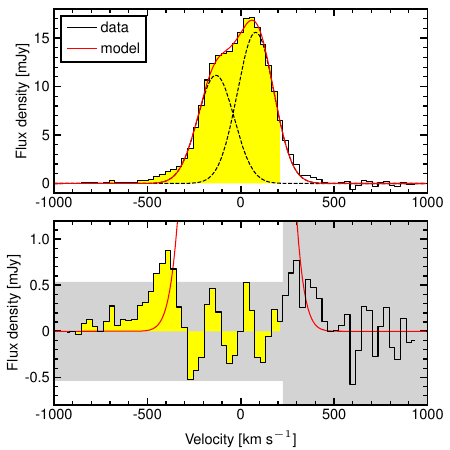}
  \caption{[C\,\textsc{ii}] spectrum in J2310+1855. Top, a black solid line shows the spectrum extracted from the 1.5\arcsec-resolution cube. One of the archival data (2019.1.01721.S) does not cover the velocity range of $v>+225$ km s$^{-1}$, which is not highlighted by yellow. A red curve represents the best-fit model, which consists of two Gaussian components shown as dashed lines, in the velocity range of $v<+225$ km s$^{-1}$. Bottom, we obtained the residual spectrum by subtracting the best-fit model from the observed spectrum. A gray-shaded region corresponds to the 5$\sigma$ level in the cube.}
  \label{fig:fig4}
\end{figure}

\begin{figure}[htbp]
  \centering
  \includegraphics[width=1.0\textwidth]{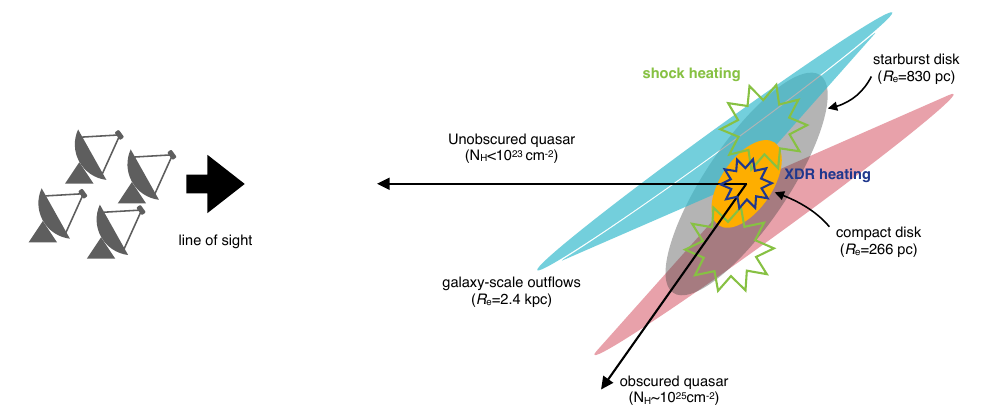}
  \caption{Schematic view of J2310+1855. This figure illustrates the composition of the quasar, which consists of a compact disk, a starburst disk, and an outflow component. It shows that XDR heating dominates as the heating mechanism in the central region of the quasar, while shock heating becomes dominant in regions slightly farther away.}
  \label{fig:fig5}
\end{figure}

\begin{figure}[htbp]
  \centering
  \includegraphics[width=0.6\textwidth]{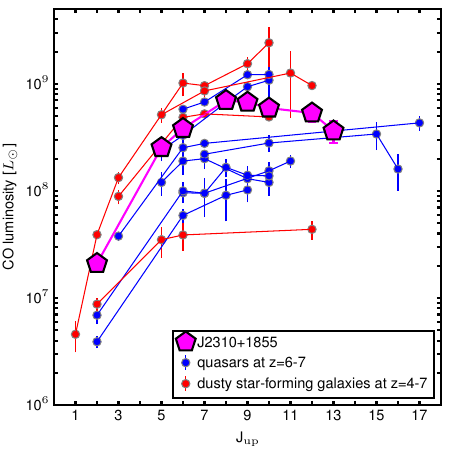}
  \caption{Galaxy-integrated CO SLEDs. A magenta pentagon, blue circles, red squares represent J2310+1855, optically luminous quasars at $z=6-7$ \cite{Gallerani2014,Li2020,Pensabene2021,Riechers2009,Wang2019,Yang2019,Li2024,Xu2024}, and dusty star-forming galaxies at $z=4-7$ \cite{Riechers2013,Salome2012,Lee2021,Jarugula2021,Tsujita2022}.}
  \label{fig:fig6}
\end{figure}

\clearpage

\section*{Methods}

\subsection*{Sample}

Quasar J2310+1855 was first identified by the Sloan Digital Sky Survey, and its spectroscopic redshift at $z=6.003$ was precisely measured by ALMA observations of [C\,\textsc{ii}] line emission \cite{Wang2013}. It is one of the brightest quasars at $z>6$, with a bolometric luminosity of $L_{\rm bol}=7.1\times10^{13}\,L_\odot$, based on an absolute magnitude at rest-frame 1450 \AA \cite{Vito2019} and a far-infrared luminosity of $L_{\rm FIR}=(1.4-1.6)\times10^{13}\,L_\odot$ \cite{Shao2019}. In addition, X-ray emission attributed to an accreting supermassive black hole (SMBH) was observed by Chandra, with a luminosity of $L_{2-10\,\mathrm{keV}}=1.8\times10^{11}\,L_\odot$ in the rest-frame 2-10 keV band \cite{Vito2019}. The mass of the black hole is calculated to be in the range of $M_\mathrm{BH}=(2-4)\times10^9\,M_\odot$, inferred from the broad Mg\,\textsc{ii} emission line \cite{Farina2022}. Figure~\ref{fig:ex-fig1}) compares the black hole mass and bolometric luminosity of J2310+1855 with those of other quasars. The Eddington ratio, which compares the bolometric luminosity to the theoretical maximum luminosity, of J2310+1855 is close to 1, suggesting that the gas accretion rate to the black hole is higher than that of most quasars discovered in this epoch. In this study, we spatially resolve the CO emission with an unprecedented high spatial resolution to investigate the properties of the interstellar medium in the vicinity of a SMBH.

\subsection*{ALMA Data}
In J2310+1855, we made new observations of CO($J=6-5$), CO($J=8-7$), CO($J=13-12$), CO($J=14-13$), and [C\textsc{ii}] line and dust continuum emissions by using ALMA (2019.1.00080.S, 2022.1.01473.S). For a [C\,\textsc{ii}] line, we additionally used dataset from the ALMA archive (2019.1.01721.S, 2018.1.00597.S, 2019.1.00661.S). The used receivers, baseline lengths, maximum recovery scale, and integration time are summarized in Figure~\ref{tab:tab1}. At $z=6.00$, both CO($J=13-12$) and CO($J=14-13$) emissions were observed simultaneously with a single frequency setup. The wavelengths of the dust continuum emissions are 3.5 mm, 3.1 mm, 2.3 mm, 2.1 mm, 1.4 mm, 1.1 mm, and 450 $\mu$m in the observed frame, corresponding to 499, 437, 328, 301, 193, 162, and 64 $\mu$m in the rest frame. Since the maximum recovery scale of the observations of CO($J=13-12$), CO($J=14-13$), and 1.4 mm continuum emissions is 0.86\arcsec (4.9 kpc), we do not discuss the total fluxes of these emissions in this paper. We use the Common Astronomy Software Application package ({\tt CASA}) \cite{CASA2022} for the data calibration. We first estimated the continuum flux densities with a first-order polynomial in the frequency range where emission lines are not included, and then subtracted them from the visibility data by using {\tt CASA/uvcontsub} task. 

For CO line emissions, we collapsed the visibility data in the velocity range from $-150$ to $+150$ km\,s$^{-1}$ and in the velocity range from $-250$ to $+250$ km\,s$^{-1}$ to increase the signal-to-noise ratio for deconvolution in imaging processing. We then adjusted the visibility weight with briggs weight to create images with different spatial resolutions from 0.028\arcsec$\times$0.018\arcsec to 0.24\arcsec$\times$0.24\arcsec by using CASA/tclean task and CASA/imsmooth task, which smooths images to an arbitrary beam. We cleaned down to the 1.5$\sigma$ noise level in a circular mask. For a [C\,\textsc{ii}] line, we created a dirty cube with a velocity width of 30 km s$^{-1}$ from all archival data. The synthesized beam size, imaging parameters (radius of circular mask, robustness parameter for Briggs weighting, and full width at half maximum of a Gaussian taper), and noise level of these images are summarized in Figure~\ref{tab:tab2}. Figure~\ref{fig:ex-fig4} and Figure~\ref{fig:ex-fig5} show the ALMA images of dust continuum emission and CO line emission, respectively. We took the peak values on the 0.08\arcsec-resolution and the 0.24\arcsec-resolution images, corresponding to the fluxes or flux densities in the central and extended regions, respectively (Figure~\ref{tab:tab3}). We then calculated the CO fluxes in the outer region at $R=230-700$ pc by subtracting those at 0.08\arcsec resolution from those at 0.24\arcsec resolution. These fluxes or flux densities are used to study CO SLEDs and dust SEDs (Fig.~\ref{fig:fig3}). 

\subsection*{Disk Modeling in the Visibility Plane}

\begin{figure}[htbp]
  \centering
  \includegraphics[width=0.6\textwidth]{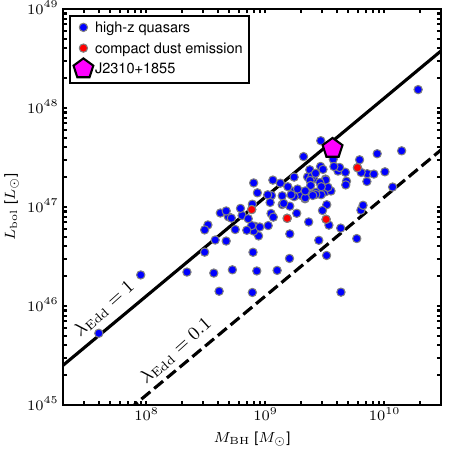}
  \caption{
  Black hole mass versus bolometric luminosity for J2310+1855 and other quasars at  $z\gtrsim5.7$ \cite{Farina2022}. Red circles indicate quasars where compact dust emission has been detected in high-resolution ALMA observations \cite{Venemans2019,Walter2022,Meyer2023,Neeleman2023}.
  }
  \label{fig:ex-fig1}
\end{figure}

\begin{figure}[htbp]
  \centering
  \includegraphics[width=1.0\textwidth]{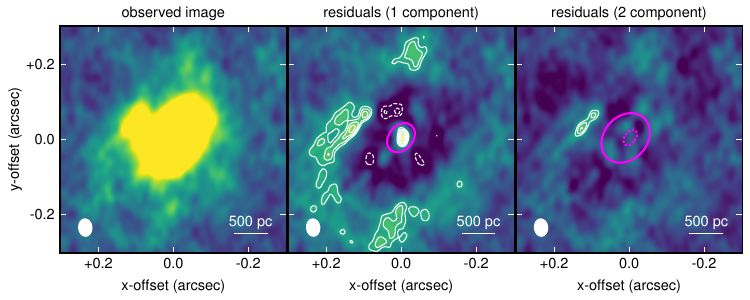}
  \caption{
  ALMA images of 1.4 mm continuum emission. From left to right, the observed clean images and the dirty images after subtraction of the best-fit one- and two-component models are shown. The spatial resolution of the images is 0.050\arcsec$\times$0.039\arcsec. 
  Dashed and solid contours are plotted every $-1\sigma$ from $-5\sigma$ and every $1\sigma$ from $5\sigma$, respectively. The effective radii of the best-fit model are shown by the magenta lines.
  }
  \label{fig:ex-fig2}
\end{figure}

Figure~\ref{fig:ex-fig2} shows that the ALMA images of the 1.4 mm continuum emission. For the disk modeling, we do not use the ALMA images, but the visibility data. The model has an exponential profile with six parameters (central position, flux density, effective radius, major-to-minor axis ratio, and position angle). We fit one- and two-component models to the ALMA data by using the {\tt UVMULTIFIT} code \cite{MartiVidal2014}. We then made dirty images after subtracting the best-fit model in the visibility plane. In the one-component model, we obtained a flux density of $S=4.98\pm0.04$ mJy, an effective radius of $R_e=0.087\arcsec\pm0.001\arcsec$, an axis ratio of $q=0.74\pm0.01$, and a position angle of PA=-41$\pm$2 degree. We found that there is residual emission above 10$\sigma$ in the center (Figure~\ref{fig:ex-fig2}). In the two-component model, we obtained $S=1.72\pm0.15$ mJy, $R_e=0.047$\arcsec$\pm$0.006\arcsec ($R_e=266\pm14$ pc), $q=0.63\pm0.03$, and PA=-35$\pm$3 degree for the compact disk, and $S=4.06\pm0.13$ mJy, $R_e=0.145$\arcsec$\pm$0.003\arcsec ($R_e=830\pm36$ pc), $q=0.79\pm0.02$, and PA=-42$\pm$4 degree for the extended disk. There is a small spatial offset of 0.012\arcsec$\pm$0.002\arcsec between the central positions of the two components. The overall emission is well characterized by the two-component model, although off-center clumps are detected at $6-7\sigma$. The compact disk with $R_e =266$ pc is roughly comparable in size to a circum-nuclear disk (CND) or ring found in nearby AGNs \cite{Izumi2013,GonzalezAlfonso2010,Combes2014}. The compact disk detected by our ALMA observations could be a CND-like component that plays a crucial role as a major reservoir of gas feeding a SMBH. The compact disk contributes about 30\% of the total dust emission. Although compact dust emission has been detected in four other quasars at $z=6-7$ \cite{Venemans2019,Walter2022,Meyer2023,Neeleman2023}, three of them do not exhibit similarly high Eddington ratios (Figure~\ref{fig:ex-fig1}). Therefore, it appears that the presence of a compact disk is not limited to phases of high gas accretion rates onto the black hole. On the other hand, the size of the extended disk with $R_e=830$ pc is comparable to that of a starburst disk/ring in nearby AGNs \cite{GonzalezAlfonso2010} (about 1 kpc).

We also modeled the wing component ($-465$ km s$^{-1}< v < -345$ km s$^{-1}$) and the disk component (-315 km s$^{-1}< v < +315$ km s-$^{-1}$) of the [C\textsc{ii}] line emission by using all the archival data. Figure~\ref{fig:ex-fig6} shows that the visibility amplitudes of [C\textsc{ii}] emission as a function of antenna separation, specifically uv distance in the visibility plane. The wing component decreases steeply at uv distances larger than 100 k$\lambda$, indicating that it is spatially extended. The failure to detect the wing component in one of the previous observations \cite{Shao2022} (2018.1.00597.S) may be due to insufficient sensitivity at short $uv$ distances ($<100$ k$\lambda$). We assumed a circular Gaussian for the wing component and an elliptical exponential disk for the disk component. The best-fit model gives $Sdv=0.170\pm0.021$ Jy km s$^{-1}$ and $R_e=0.42\arcsec\pm0.07\arcsec$ for the wing component (Figure~\ref{fig:ex-fig6}) and $Sdv=7.85\pm0.05$ Jy km s$^{-1}$, $R_e=0.28\arcsec\pm0.01\arcsec$, $q=0.70\pm0.02$, and PA=322$\pm$2 degree for the disk component. The circularized effective radius of the disk component is $R_{e,circ} =0.23\arcsec\pm0.01\arcsec$. This value is slightly larger than that reported in a previous image-based analysis ($R_{e,circ} =0.19$\arcsec 26; 2019.1.00661.S and 2019.1.01721.S), but is likely due to differences in measurement methods and datasets. We derived $R_{e,circ} =0.20\arcsec\pm0.01\arcsec$ by fitting only the same visibility data from 2018.1.00597.S to an elliptical exponential disk model.

\subsection*{PDR and XDR Models}

We used the {\tt galaxySLED} code to produce model CO SLEDs \cite{Esposito2024}. While many theoretical models assume a constant gas density, our model considers giant molecular clouds (GMCs), which are composed of molecular clumps with different gas densities. The clump density distribution within a GMC follows a lognormal distribution, and 15 types of GMCs with different masses and sizes are available in galaxySLED. We distributed the 15 GMCs to create an exponential disk model with a scale length of $R_d =R_e /1.68$ and a total gas mass of $M_\mathrm{gas}$. We assumed a thick disk with a ratio of scale height to radial scale length of $q=z_d/R_d=0.3$ \cite{Sani2012}.

\begin{figure}[htbp]
  \centering
  \includegraphics[width=0.6\textwidth]{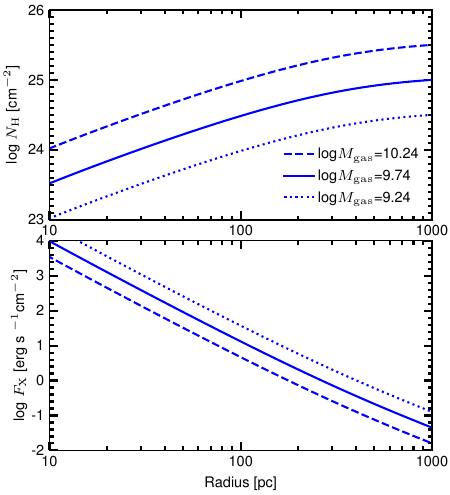}
  \caption{
  XDR model. A gas column density (top) and X-ray flux (bottom) at different galactocentric radii. The dotted, solid, and dashed lines show the XDR models with different gas masses.
  }
  \label{fig:ex-fig3}
\end{figure}

We consider a situation where there is an AGN with X-ray luminosity $L_X$ at the center of the gas disk. The incident X-ray flux $F_X (r)$ at a distance r from the center depends on the radial gas column density, since it is attenuated by the gas in the disk. Therefore, in a gas disk characterised by three parameters $M_\mathrm{gas},R_d ,q$, we calculated the average radial column density as follows.

\[
N_{\mathrm{H}}(r)
= \int_{0}^{r} \frac{M(x)}{\mu m_{p}\,V(x)} \, dx
= \frac{2.08 \times 10^{4} \, M_{\mathrm{gas}}}{\mu \, m_{p}}
  \int_{0}^{r} \frac{1 - \exp\!\bigl(-x/R_{d}\bigr)\,\bigl(\tfrac{x}{R_{d}} + 1\bigr)}%
  {2\pi \, q \, R_{d} \, x^{2}} \, dx,
\]

\noindent
where $\mu$ is the mean molecular weight and $m_p$ is the mass of a proton. In J2310+1855, we know that $R_e=266$ pc from the ALMA observations of 1.4 mm continuum emission and $L_{2-10\,\mathrm{keV}}=6.9\times 10^{44}$ erg s$^{-1}$ from X-ray observations \cite{Vito2019}. 
Although $M_\mathrm{gas}$ is the only free parameter in our model, the fundamentally important physical parameter contributing to X-ray attenuation is the gas column density. Because the gas mass and the disk thickness are degenerate for a given gas column density, we do not use the gas mass for discussion in this work. 
In Fig.~\ref{fig:ex-fig3}, we show the gas column densities and X-ray fluxes at different galactocentric radii in three models with different gas masses of $M_\mathrm{gas}=10^{9.24,9.74,10.24}~M_\odot$. In these models, the gas column densities reach $N_H(r=1 \mathrm{kpc})=10^{24.5, 25.0, 25.5}$ cm$^{-2}$ at a galactocentric radius of 1 kpc. 
In each GMC distributed at different galactocentric radii in the gas disk, the CO fluxes are calculated using the photoionization code {\tt CLOUDY} \cite{Ferland2017} under different incident X-ray flux conditions. 
The temperature of the warm gas in J2310+1855 is estimated to be above 200 K \cite{Li2020}. At $z=6$, the Cosmic Microwave Background (CMB) temperature is 18 K, which significantly affects the excitation states of cold gas below 100 K \cite{daCunha2013}. 
However, the CMB should have little impact on warm gas above 200 K. As the gas column density in the radial direction of the disk increases, CO becomes less excited. In the PDR model, the incident UV flux does not depend on the galactocentric radius because star formation is assumed to occur throughout the gas disk. 
The integrated CO flux from these GMCs is used in Fig.~\ref{fig:fig2}. In the right panel of Fig.~\ref{fig:fig3}, we normalized the model CO SLEDs to the CO $J=13-12$ line, as this avoids the need to rely on a CO-to-H$_2$ conversion factor.

\subsection*{Dust SED}

\begin{figure}[htbp]
  \centering
  \includegraphics[width=1.0\textwidth]{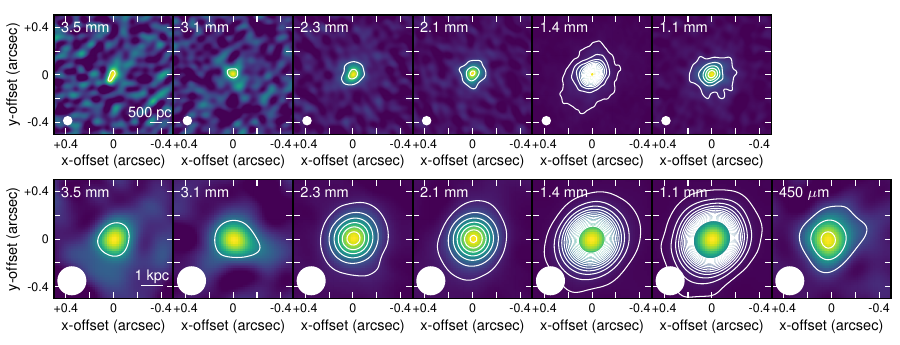}
  \caption{
  ALMA image of dust continuum emission. The beam size is 0.08\arcsec$\times$0.08\arcsec~(top) and 0.24\arcsec$\times$0.24\arcsec~(bottom). The contours are plotted every $10\sigma$ from $5\sigma$.
  }
  \label{fig:ex-fig4}
\end{figure}

\begin{figure}[htbp]
  \centering
  \includegraphics[width=1.0\textwidth]{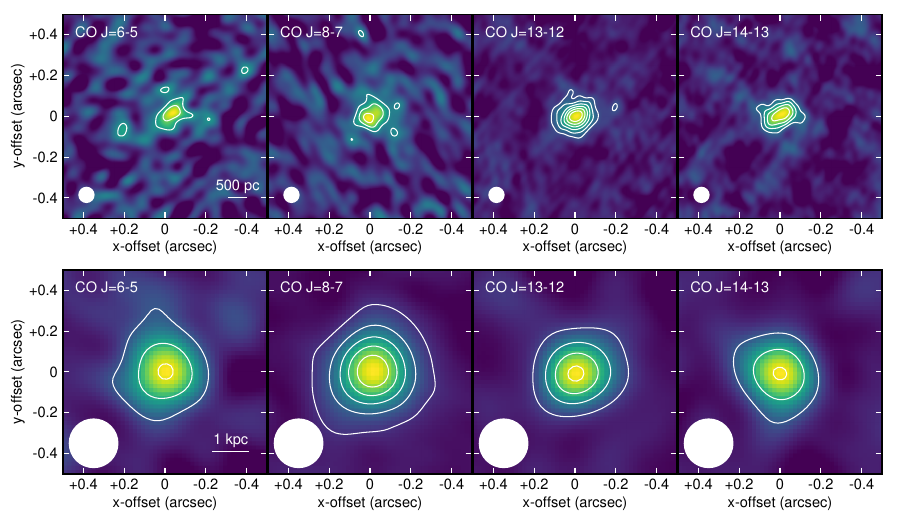}
  \caption{
  ALMA image of CO emission. The beam size is 0.08\arcsec$\times$0.08\arcsec~(top) and 0.24\arcsec$\times$0.24\arcsec~(bottom). The contours are plotted every $2\sigma$ from $3\sigma$.
  }
  \label{fig:ex-fig5}
\end{figure}

\begin{figure}[htbp]
  \centering
  \includegraphics[width=0.6\textwidth]{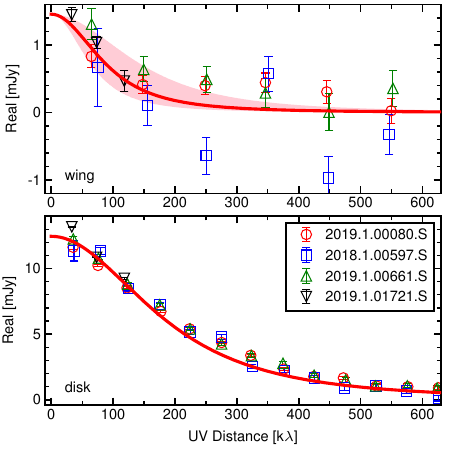}
  \caption{
  Visibility amplitude of [C\,\textsc{ii}] emission. Top, a wing component in in the velocity range of $-465<v<-345$ km\,s$^{-1}$. Bottom, a disk component in in the velocity range of $-315<v<+315$ km\,s$^{-1}$. 
  Different symbols show different observations (Fig.~\ref{tab:tab1}). We show the real part of the visibility as a function of uv distance along the minor axis, corresponding to the major axis in the images. A red line and red shaded region show the best-fit model and its uncertainties, respectively.
  }
  \label{fig:ex-fig6}
\end{figure}

\begin{figure}[htbp]
  \centering
  \includegraphics[width=1.0\textwidth]{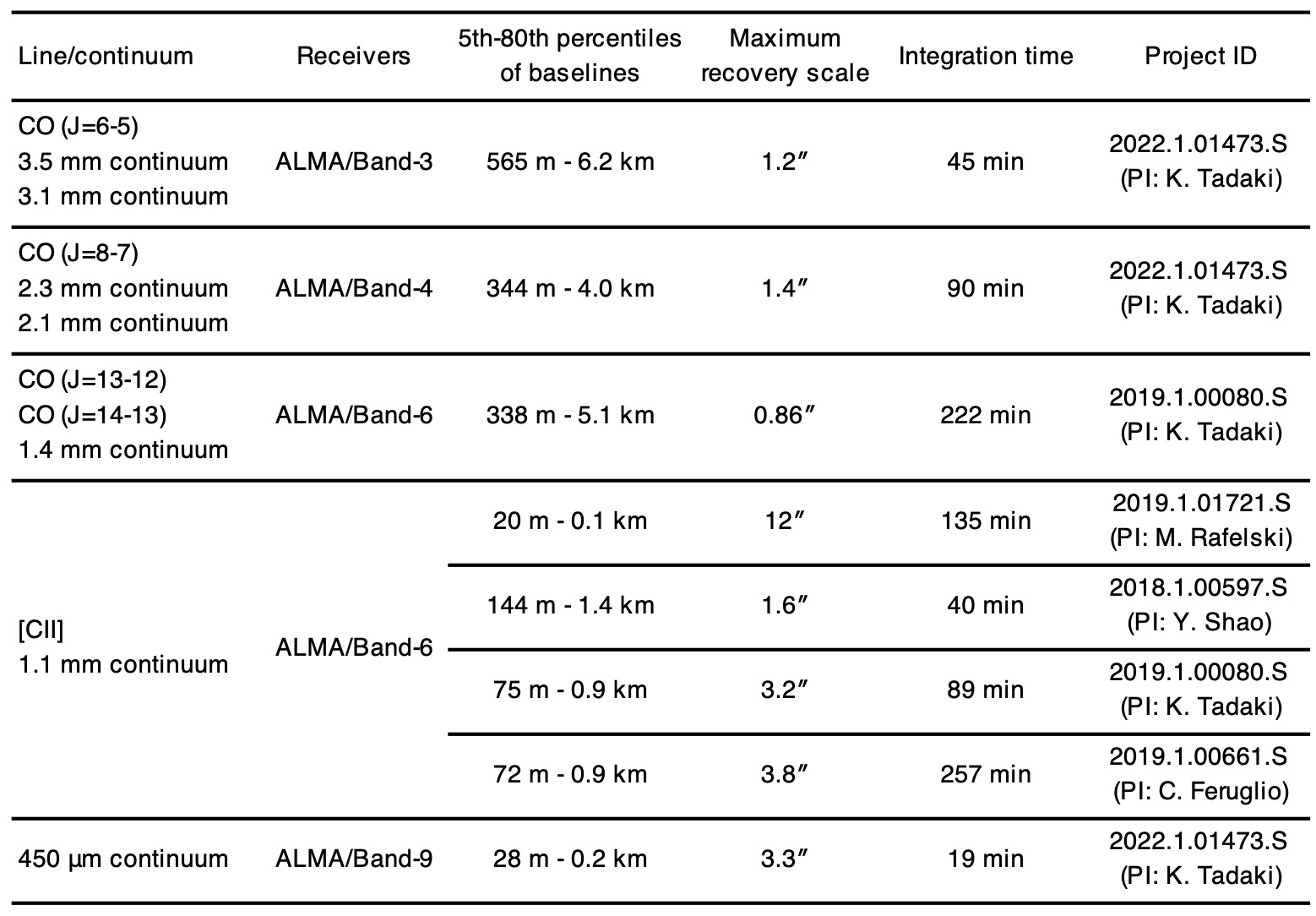}
  \caption{Summary of ALMA observations.}
  \label{tab:tab1}
\end{figure}

\begin{figure}[htbp]
  \centering
  \includegraphics[width=1.0\textwidth]{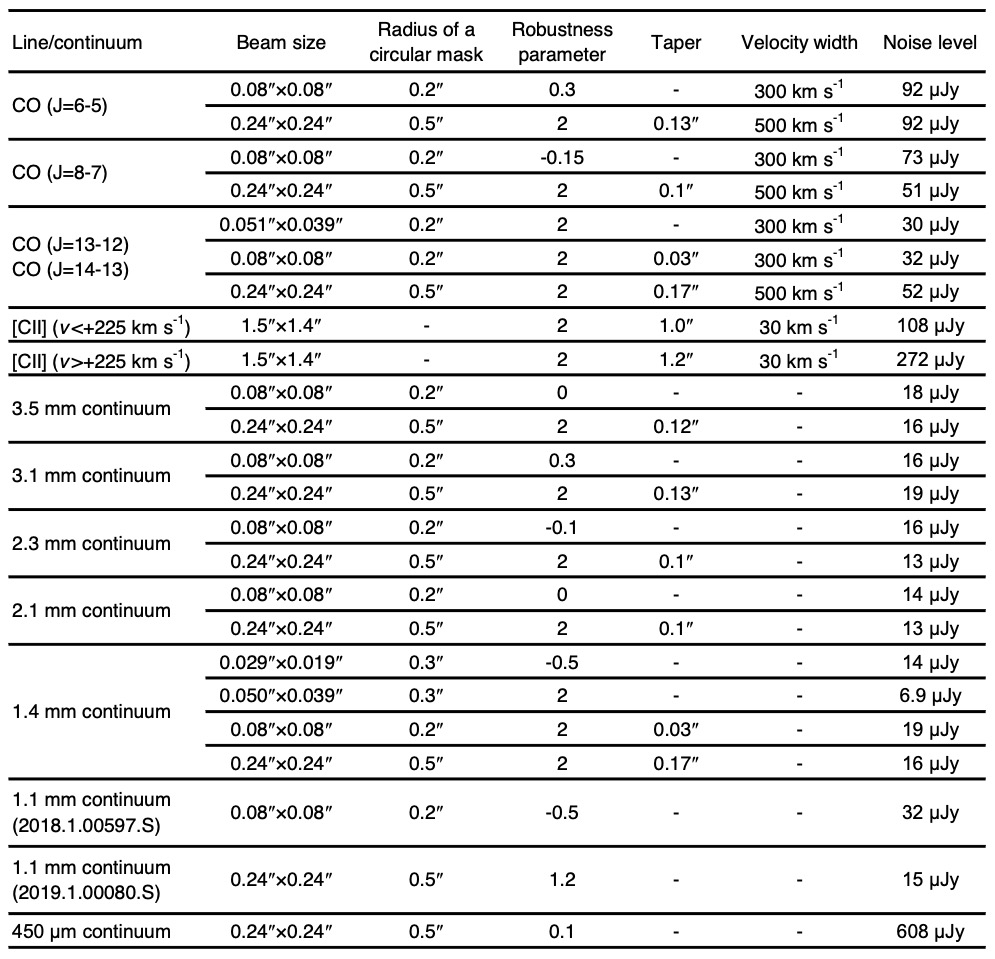}
  \caption{Summary of image processing.}
  \label{tab:tab2}
\end{figure}

\begin{figure}[htbp]
  \centering
  \includegraphics[width=0.8\textwidth]{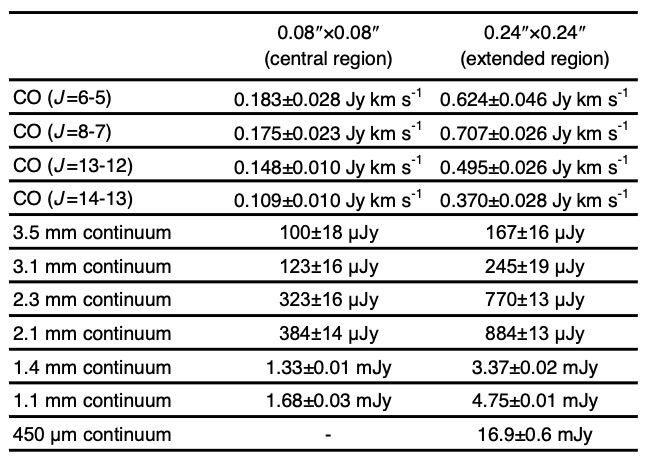}
  \caption{Peak flux densities in the ALMA images.}
  \label{tab:tab3}
\end{figure}

We detected multi-wavelength continuum emissions at resolutions of 0.08\arcsec and 0.24\arcsec (Fig.~\ref{fig:ex-fig4} and Fig.~\ref{tab:tab3}), providing the dust SED in the central region and extended regions, respectively. We used the {\tt MERCURIUS} code \cite{Witstok2022} to fit a modified blackbody radiation model to the ALMA dust emission data. 
The model has four parameters of dust mass $M_d$ (or a total infrared luminosity $L_\mathrm{IR}$), dust temperature $T_d$ and an emissivity index $\beta$, the wavelength at with the dust optical depth is unity $\lambda_0$. 
We adopted a self-consistent opacity model where the optical depth is proportional to the surface density of the dust mass. We assumed that dust emission is homogeneously distributed in the synthesized beam and that the area of dust emission is equal to the effective area of the beam (0.236 kpc$^2$ and 2.20 kpc$^2$). 450 $\mu$m (rest-frame 64 $\mu$m) photometry in the Band-9 is available only for the SED in the extended region. In addition to the measurement uncertainties, the flux calibration uncertainties of 5\% in Band-3 and -4, 10\% in Band-6, and 20\% in Band-9 are included in the fitting. 
The best-fit model gives $M_d=3.1^{+1.1}_{-0.9}\times10^8\,M_\odot$, $L_{\rm IR}=3.8^{+1.7}_{-1.0}\times10^{12}\,L_\odot$, $T_d=74^{+7}_{-5}$ K, $\beta=1.7\pm0.2$, and $\lambda_0=266^{+52}_{-49}\,\mu$m in the central region and $M_d=7.0^{+1.7}_{-1.3}\times10^8\,M_\odot$, $L_{\rm IR}=1.2^{+0.4}_{-0.3}\times10^{13}\,L_\odot$, $T_d=59\pm4$ K, $\beta=1.6\pm0.1$, and $\lambda_0=115^{+16}_{-13}\,\mu$m in the extended region. 
Previous studies of spatially unresolved dust SED including Herschel photometry ($100-500~\mu$m) show average dust temperatures of $T_d$=53 K \cite{Shao2022} and $T_d$=71 K \cite{Tripodi2022}. 
In galaxy-integrated SEDs at $100-350~\mu$m, the contribution of an AGN torus becomes significant. The dust temperature of a cold component is affected by the choice of torus model. In addition, by averaging the dust emission over the entire galaxy, the surface density of the dust mass becomes smaller compared to that in the central region of the galaxy, thus reducing the apparent optical depth. Therefore, it is difficult to directly compare the results of resolved and unresolved dust SEDs.

\subsection*{Dust Correction for CO Line Emission}

The correction for dust attenuation of the CO line emission is based on the literature \cite{Rangwala2011}. This approach assumes that gas and dust are well mixed. As the optical depth of the dust increases, the observed CO flux ($F$) is attenuated from its intrinsic flux ($F_0$) according to $F = F_0 \frac{1-e^{-\tau_\lambda}}{\tau_\lambda}$. For observations of low-$J$ CO line emission in normal star-forming galaxies, the dust is optically thin at these wavelengths. However, for observations of high-J CO line in extremely dusty objects such as Arp 220, the optical depth exceeds unity at these wavelengths, requiring an attenuation correction. 
From the dust SED of J2310+1855 shown in the left panel of Fig.~\ref{fig:fig3}, we find that the wavelength at which the dust optical depth reaches unity is $\lambda_0=266\,\mu$m and the wavelength dependence of the optical depth is $\tau_\lambda=(\lambda/\lambda_0)^{-1.7}$. 
Therefore, the correction factor ($F_0/F$) becomes 1.2, 1.4, 2.0, and 2.2 at the wavelengths of the CO $J=6-5$, $J=8-7$, $J=13-12$, and $J=14-13$ line, respectively.

\clearpage
\section*{Extended Data}

\clearpage

\section*{Acknowledgements}

\begin{sloppypar}
This paper makes use of the following ALMA data: ADS/JAO.ALMA\#2018.1.00597.S, ADS/JAO.ALMA\#2019.1.00080.S, ADS/JAO.ALMA\#2019.1.00661.S, ADS/JAO.ALMA\#2019.1.01721.S, ADS/JAO.ALMA\#2022.1.01473.S. ALMA is a partnership of ESO (representing its member states), NSF (USA) and NINS (Japan), together with NRC (Canada), NSTC and ASIAA (Taiwan), and KASI (Republic of Korea), in cooperation with the Republic of Chile. The Joint ALMA Observatory is operated by ESO, AUI/NRAO and NAOJ. 
K.T. acknowledges support from JSPS KAKENHI Grant Number JP 23K03466 and 23K20870. F.E. acknowledges support from grant PRIN MIUR 2017-20173ML3WW\_s, and funding from the INAF Mini Grant 2022 program "Face-to-Face with the Local Universe: ISM's Empowerment (LOCAL)". 
T.M. is supported by a University Research Support Grant from the National Astronomical Observatory of Japan (NAOJ). Data analysis was in part carried out on the Multi-wavelength Data Analysis System operated by the Astronomy Data Center (ADC), National Astronomical Observatory of Japan.
\end{sloppypar}


\bibliographystyle{plain}
\bibliography{references}

\end{document}